# VW-PINNs: A volume weighting method for PDE residuals in physics-informed neural networks

Jiahao Song[1], Wenbo Cao[1], Fei Liao[1], Weiwei Zhang[1,*]

[1] *School of Aeronautics, Northwestern Polytechnical University, Xi'an 710072, China*

**Abstract**

Physics-informed neural networks (PINNs) have shown remarkable prospects in the solving the forward and inverse problems involving partial differential equations (PDEs). The method embeds PDEs into the neural network by calculating PDE loss at a series of collocation points, providing advantages such as meshfree and more convenient adaptive sampling. However, when solving PDEs using nonuniform collocation points, PINNs still face challenge regarding inefficient convergence of PDE residuals or even failure. In this work, we first analyze the ill-conditioning of the PDE loss in PINNs under nonuniform collocation points. To address the issue, we define volume-weighted residual and propose volume-weighted physics-informed neural networks (VW-PINNs). Through weighting the PDE residuals by the volume that the collocation points occupy within the computational domain, we embed explicitly the spatial distribution characteristics of collocation points in the residual evaluation. The fast and sufficient convergence of the PDE residuals for the problems involving nonuniform collocation points is guaranteed. Considering the meshfree characteristics of VW-PINNs, we also develop a volume approximation algorithm based on kernel density estimation to calculate the volume of the collocation points. We verify the universality of VW-PINNs by solving the forward problems involving flow over a circular cylinder and flow over the NACA0012 airfoil under different inflow conditions, where conventional PINNs fail; By solving the Burgers' equation, we verify that VW-PINNs can enhance the efficiency of existing the adaptive sampling method in solving the forward problem by 3 times, and can reduce the relative error of conventional PINNs in solving the inverse problem by more than one order of magnitude.

**Keywords:** Physics-informed neural networks, Partial differential equations, Nonuniform sampling, Computational physics, Deep learning

## 1 Intruduction

In recent years, physics-informed neural networks (PINNs) [1] have became a research hotspot for solving the forward and inverse problems involving partial differential equations (PDEs). The central idea of PINNs is integrating the governing equations of physical systems into the loss function of neural networks, ensuring that the neural networks minimize PDE residuals while approaching the definite conditions or observed data. In fact, the concept of PINNs can be traced back to the 20th century when Dissanayake et al. [2] pioneered the use of neural networks to solve partial differential equations. Compared with mesh-based numerical methods such as finite element (FEM), finite difference (FDM), and finite volume (FVM), the advantage of PINNs lies in their ability to directly calculate spatiotemporal derivatives through automatic differentiation (AD) [3], enabling a meshfree approach that avoids mesh generation, which also indicates that PINNs can alleviate the curse of dimensionality to a certain extent [4-5]. Moreover, PINNs can conveniently integrate various observed data for solving the inverse problems, such as inferring velocity and pressure fields based on concentration field and the Navier-Stokes equations [6]. In contrast, numerical methods demand extremely high cost to achieve the same process. PINNs have been demonstrated for various physics phenomena, including fluid mechanics [6-9], heat transfer [10-11], fluid-structure interaction [12-13], electromagnetic propagation [14], and quantum chemistry [15].

Although PINNs have achieved gratifying performance, it still faces challenges in terms of accuracy, computational efficiency and training robustness, especially for complex problems [16]. Therefore, over the past few years, numerous researchers have enhanced PINNs in various aspects. For example, adaptive activation functions have been designed to improve the convergence of the algorithm [17], and *hp-VPINNs* based on the weak form of PDE have been proposed to perform integral transformation on the PDE residuals through Legendre basis function [18]. gPINNs have been developed to embed gradient information of PDE residuals into the loss function [19]. For complex spatiotemporal problems, the parallel frameworks [20-21] have been built based on spatiotemporal domain decomposition to accelerate the training of PINNs and improve the accuracy. In order to enhance multi-scale recognition capabilities, Fourier feature network [22] that scales spatio-temporal coordinates in sinusoidal space and MscaleDNN [23] that constructs multi-scale input a priori were developed. Furthermore, the gradient pathologies in PINNs have been addressed through adaptive weight [24-26] and hard constraints [27-29].

In PINNs, PDE loss is evaluated at a set of scattered collocation points. The effect of collocation points on PINNs is similar to the effect of mesh points on FEM.

Thus, the location and distribution of these collocation points should be highly important to the performance of PINNs [30]. When the solution of the PDE is a simple continuous smooth function, uniform sampling of the collocation points is usually appropriate. However, for many practical problems, such as flow over an object and shock wave capture in fluid mechanics. The gradients of the solution at the object and shock wave are large, while they are small in other computational regions. Thus, refining the collocation points near the object and shock wave to better capture the details of flow while maintaining low-density sampling in other regions can consider both computational cost and accuracy, which is already the consensus for solving such problems by numerical methods. At present, although some researches on PINNs have paid attention to the importance of nonuniform sampling [30-39], these studies still calculate the mean squared error of PDE residuals at all collocation points as PDE loss. This loss evaluation method overlooks the inconsistent convergence of PDE residuals at different locations within the computational domain caused by differences in sampling density. Specifically, in the PDE loss of PINNs, the proportion of PDE residuals at all collocation points is equal. The reduction of local PDE residuals within high-density regions brings greater benefits of loss reduction compared with those within low-density regions when solving problems using nonuniform collocation points. The network naturally focuses on reducing the PDE residuals in these high-density sampling regions. Conversely, the residuals in low-density sampling regions have difficulty converging. Consequently, it diminishes the efficiency of PINNs and may even lead to failing, especially when there is a significant difference in sampling density. We refer to this issue as the ill-conditioning of PDE loss function.

To address the issue, this work defines volume-weighted residual and proposes volume-weighted physics-informed neural networks (VW-PINNs). The PDE loss function is re-evaluated by weighting the PDE residuals based on the volume occupied by the collocation points within the computational domain. To calculate the volume of collocation points in meshfree scenarios such as VW-PINNs, we also develop an efficient volume approximation algorithm based on kernel density estimation [40-41]. By solving four forward problems and one inverse problem, we verify the advantages of the proposed method in terms of universality, convergence, and accuracy.

The remainder of the paper is organized as follows. In Section 2, we provide a detailed introduction to PINNs and analyze the ill-conditioning of the PDE loss function in PINNs. Then, we propose VW-PINNs. In Section 3, We carry out a series of numerical experiments to verify the effectiveness of VW-PINNs. Concluding remarks and direction for future research are then presented in Section 4.

## 2 Methodology

2.1 Physics-informed neural networks（PINNs）

Differing from numerical methods, PINNs transform solving PDEs into an optimization problem. Its input usually involves spatiotemporal coordinates $(t, \mathbf{x})$, corresponding to the spatiotemporal variables of the PDEs, with the output being the solution $u(t, \mathbf{x})$. We consider the PDE defined on a domain $\Omega \in \mathbb{R}^d$:

$$\begin{cases} u_t + L[u] + N[u] = 0, & t \in [0,T], \mathbf{x} = (x_1,...,x_d) \in \Omega \\ u(0,\mathbf{x}) = g(\mathbf{x}), & \mathbf{x} \in \Omega \\ B[u] = 0, & t \in [0,T], \mathbf{x} \in \partial\Omega \end{cases} \quad (1)$$

where $u$ denotes the solution of the PDE, $L[\cdot]$ represents the linear differential operator, $N[\cdot]$ represents the nonlinear differential operator, and $B[\cdot]$ represents the boundary condition operator.

Regarding this problem, the loss function of the PINNs is defined as follows:

$$L = \omega_{ic} L_{ic} + \omega_{bc} L_{bc} + \omega_r L_r + \omega_{da} L_{da} \quad (2)$$

where

$$L_{ic} = \frac{1}{N_{ic}} \sum_{i=1}^{N_{ic}} \left| u_\theta(\mathbf{x}_{ic}^i, 0) - g(\mathbf{x}_{ic}^i) \right|^2 \quad (3)$$

$$L_{bc} = \frac{1}{N_{bc}} \sum_{i=1}^{N_{bc}} \left| B[u_\theta(\mathbf{x}_{bc}^i, t_{bc}^i)] \right|^2 \quad (4)$$

$$L_r = \frac{1}{N_r} \sum_{i=1}^{N_r} \left| \frac{\partial u_\theta}{\partial t}(\mathbf{x}_r^i, t_r^i) + L[u_\theta(\mathbf{x}_r^i, t_r^i)] + N[u_\theta(\mathbf{x}_r^i, t_r^i)] \right|^2 \quad (5)$$

$$L_{da} = \frac{1}{N_{da}} \sum_{i=1}^{N_{da}} \left| u_\theta(\mathbf{x}_{da}^i, t_{da}^i) - u(\mathbf{x}_{da}^i, t_{da}^i) \right|^2 \quad (6)$$

In formulas (3)-(6), $\theta$ represents the network parameters, while $u_\theta(\mathbf{x}, t)$ denotes the solution of PINNs. $\{\mathbf{x}_{ic}^i, 0\}_{i=1}^{N_{ic}}$, $\{\mathbf{x}_{bc}^i, t_{bc}^i\}_{i=1}^{N_{bc}}$, and $\{\mathbf{x}_r^i, t_r^i\}_{i=1}^{N_r}$ are training points for the initial conditions, boundary conditions, and PDE residuals, typically sampled directly within the computational domain. $u(\mathbf{x}_{da}^i, t_{da}^i)$ represents available observed data obtained through experiment or other methods, and the corresponding loss function (6) only exists in solving the inverse problems. During the network training, $L_{ic}$, $L_{bc}$, and $L_{da}$ ensure the network outputs approach the initial conditions, boundary conditions and observed data, while $L_r$ constrains them to satisfy the governing equation. The conventional PINNs framework is illustrated in Figure 1.

The weight parameters, $\omega_{ic}$, $\omega_{bc}$, $\omega_r$ and $\omega_{da}$, control the balance between different components in the loss function. The appropriate weight can speed up the convergence rate of PINNs training [42]. Due to the requirement for computing high-order derivatives in PINNs, the activation functions should be highly differentiable, with commonly used options including hyperbolic tangent function (tanh) and sine function (sin). Optimization algorithms usually employ Adam [43] and L-BFGS [44].

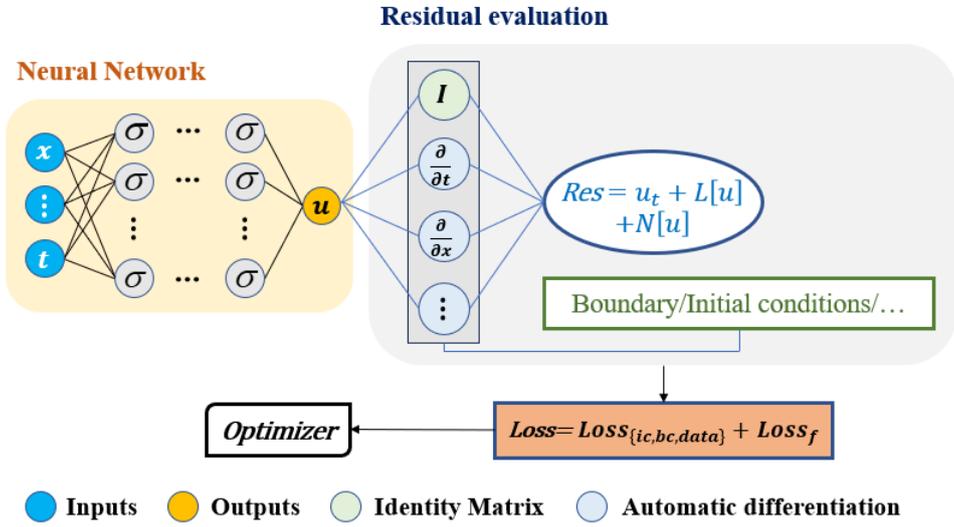

Figure 1. A Schematic of PINNs for solving partial differential equations.

## 2.2 The ill-conditioning of PDE loss function in PINNs

At present, most researches on PINNs generally compute the PDE loss according to formula (5), which means that the proportion of PDE residuals at all collocation points is equal in the loss. It is reasonable when employing uniform sampling for collocation points, because the spatial distribution pattern of the collocation points is consistent with their proportion in the loss. However, when employing nonuniform sampling, due to the sampling density disparities among collocation points at different locations in the computational domain, computing the loss according to formula (5) will lead to the domination of PDE residuals from high-density sampling regions in the network training. The network parameters tend to sufficiently reduce the residuals in these regions, in contrast, other low-density regions receive little attention and the residuals converge difficultly. In other words, the decrease of conventional PDE loss can't guarantee the convergence of the total PDE residual in the entire computational domain. The reason is as follows: considering the high-density and low-density sampling regions that occupy the same volume within the computational domain, it is evident that the former contains a greater number of collocation points than the latter. Thus, when the PDE residuals at all collocation points have an equal proportion in the PDE loss, the loss reduction benefits brought by reducing the local residuals in high-density regions are higher than those in low-density regions. Inevitably, the networks naturally focus on local learning in these high-density regions. But as is well known, solving partial differential equations requires an adequate reduction in the residuals throughout the entire computational domain. As according to the error propagation theory, erroneous results in regions with high residuals can lead to inaccuracies in the solution of other regions, even if the residuals in those regions are low [38]. Therefore, for nonuniform sampling problems, conventional PDE loss exhibits obvious limitations, it diminishes the solving efficiency of PINNs and may even lead to solving failure, especially for complex nonlinear problems with significant differences in sampling density.

We take the flow over an object in fluid mechanics as an example to illustrate the above issue, which is a typical nonuniform sampling problem. Because the gradient of the flow field is large near the object, a sufficiently dense set of collocation points is required to capture the flow details better. In contrast, the gradient in other regions is smaller, and sparse collocation points are adequate, which is the consensus for solving such problems. However, current researches using PINNs to solve the forward problems involving flow over an object commonly relies on nearly uniform collocation points within the channel [45-47] or within a specific local region [7,48]. In these cases, PDEs are solved within small computational domains, which are only several times the size of the objects. Thus, it is acceptable to use dense collocation points in regions far from the object. However, for classical external flow problems at subsonic or transonic speeds, the far-field boundary is infinite in principle and is at least dozens of times larger than the object in practice. In such scenarios, uniformly sampling according to the resolution requirement near the object would result in an enormous number of collocation points, which leads to significantly high computational cost in solving problems. Therefore, nonuniform sampling is necessary. We use PINNs to solve the inviscid compressible flow over a circular cylinder and the collocation points obtained by nonuniform sampling. The governing equations for this problem are:

$$\rho(\frac{\partial u}{\partial x}+\frac{\partial v}{\partial y})+u\frac{\partial \rho}{\partial x}+v\frac{\partial \rho}{\partial y}=0$$
$$\rho(u\frac{\partial u}{\partial x}+v\frac{\partial u}{\partial y})+\frac{\partial p}{\partial x}=0$$
$$\rho(u\frac{\partial v}{\partial x}+v\frac{\partial v}{\partial y})+\frac{\partial p}{\partial y}=0 \quad (7)$$
$$\rho c_v(u\frac{\partial T}{\partial x}+v\frac{\partial T}{\partial y})+\gamma^2 p(\frac{\partial u}{\partial x}+\frac{\partial v}{\partial y})=0$$

In formula (7), $u$ denotes the $x$-component of the velocity field, $v$ the $y$-component. $\rho$, $T$ and $p$ represent density, temperature and pressure respectively, satisfying the relationship: $p = \rho T/(\gamma Ma^2)$ ($Ma$ is the Mach number). $c_v$ is the specific heat at constant volume, $\gamma = 1.4$ is the specific heat ratio, satisfying the relationship: $c_v = \gamma/[(\gamma-1)Ma^2]$. We set $Ma = 0.4$, with $[x, y]$ as the network inputs and $[\rho, u, v, T]$ as the network outputs.

The cylinder is placed at $(x, y) = (0,0)$ with diameter $D = 1$, while the far field is positioned at the same center with a diameter of 40. The computational domain and collocation points distribution are shown in Figure 2. The total number of collocation points is $N_r = 4800$. Meanwhile, we randomly select $N_{bc} = 80$ points on the cylinder and apply the non-penetrating condition $V \cdot n = 0$. In the far field, we similarly choose $N_{bc} = 80$ points randomly and specify the dimensionless variables as constants $[\rho_\infty, u_\infty, v_\infty, T_\infty] = [1,1,0,1]$ (Because the flow away from the object approaches the uniform freestream). We solve this problem using conventional PINNs with 5 hidden layers and 64 neurons per layer, the activation function chosen as tanh. For the network training, we first run 3000 steps using the Adam optimizer, followed by an additional 2000 steps using the L-BFGS optimizer (maximum number of inner iterations is 20).

Figure 3 displays the convergence of loss functions obtained by PINNs. Evidently, both the boundary loss and the PDE loss are decreased by more than 5 orders of magnitude, which usually means that the networks obtain acceptable results. However, the flow field given by PINNs are wrong, as shown in Figure 4. The distribution of PDE residuals shown in Figure 5 visually illustrates the reason for the solving failure. While the residuals near the cylinder with high-density sampling converge to the order of $10^{-4}$, the residuals in the far-field freestream region with low-density sampling only drop to the order of $10^{-1}$ (Although the residuals in the far-field freestream region are large, the averaged PDE loss remains low due to the small number of collocation points in this region). Which indicates that the total PDE residual across the entire computational domain insufficiently converges. The information from the far-field freestream fails to propagate to the vicinity of the cylinder, resulting in an inevitable failure of the solving. Our perspective receives strong validation from this result.

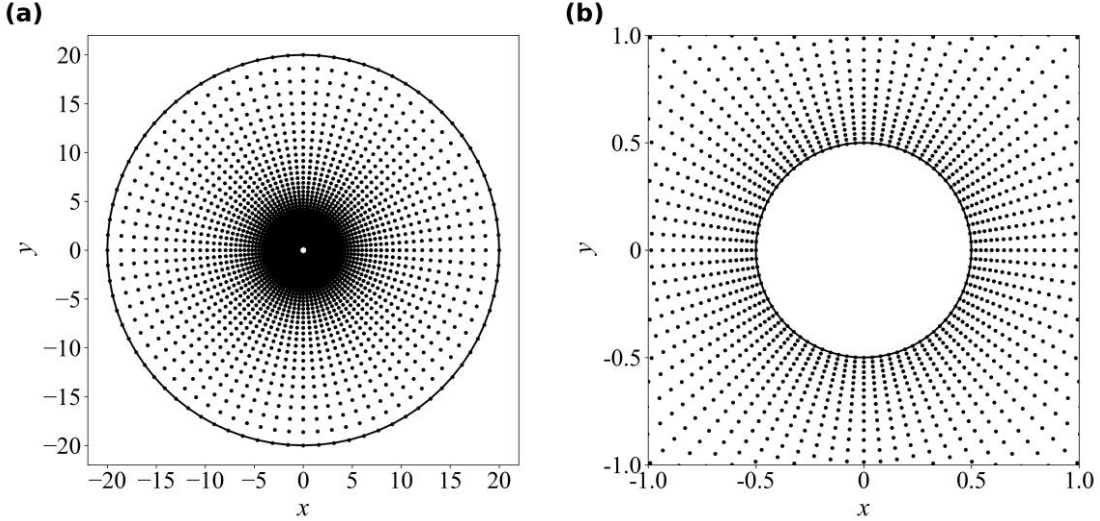

Figure 2. The computational domain and the distribution of collocation points in solving inviscid compressible flow over a circular cylinder. (a) Entire computational domain. (b) Near the cylinder region.

2.3 Volume-weighted physics-informed neural networks (VW-PINNs)

To address the issue described in Section 2.2, based on the volume occupied by collocation points within the computational domain, we define the volume-weighted residual. Taking partial differential equation (1) as an example, the volume-weighted residual at collocation point $(\mathbf{x}_r^i, t_r^i)$ is:

$$Res_v = \{\frac{\partial u_\theta}{\partial t}(\mathbf{x}_r^i, t_r^i) + L[u_\theta(\mathbf{x}_r^i, t_r^i)] + N[u_\theta(\mathbf{x}_r^i, t_r^i)]\}V(\mathbf{x}_r^i, t_r^i) \qquad (8)$$

where $V(\mathbf{x}_r^i, t_r^i)$ is the volume occupied by $(\mathbf{x}_r^i, t_r^i)$ in the computational domain. Compared with the conventional calculation of only the residual of PDE at $(\mathbf{x}_r^i, t_r^i)$, we weight the residuals of PDE through the volume occupied by collocation points in the computational domain, embedding explicitly the spatial distribution characteristics of collocation points in the residual evaluation. Building on this, we propose volume-weighted physics-informed neural networks (VW-PINNs), establishing a novel method for evaluating the PDE loss function. Still taking the partial differential equation (1) as an example, the PDE loss of VW-PINNs is:

$$L_r = \sum_{i=1}^{N_r} \left| \{\frac{\partial u_\theta}{\partial t}(\mathbf{x}_r^i, t_r^i) + L[u_\theta(\mathbf{x}_r^i, t_r^i)] + N[u_\theta(\mathbf{x}_r^i, t_r^i)]\} V(\mathbf{x}_r^i, t_r^i) \right|^2 / \sum_{i=1}^{N_r} [V(\mathbf{x}_r^i, t_r^i)]^2 \quad (9)$$

Compared with the PDE loss of conventional PINNs, VW-PINNs impose PDE constraint through volume-weighted residual. The residual of PDE at each collocation point is weighted by the volume it occupies, and the weight satisfies a linear inverse relationship with the sampling density of collocation point. Therefore, our method balances the differences in residual convergence across regions with various sampling densities in network optimization. In other words, the decrease of volume-weighted PDE loss can ensure effective convergence of the total PDE residual across the entire computational domain. The convergence level of the volume-weighted PDE loss determines the convergence status of the total residual of PDE. In addition, it is obvious that when uniform sampling is employed, $V(\mathbf{x}_r^i, t_r^i) = V_s / N_r$ ($V_s$ is the volume of the entire computational domain), formula (9) degenerates into formula (5).

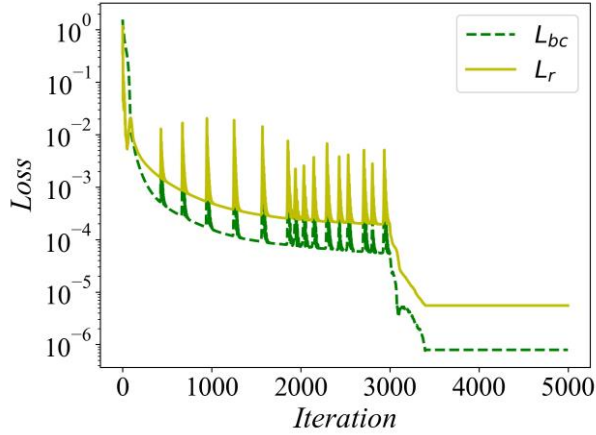

Figure 3. The convergence of loss functions obtained by PINNs for solving inviscid compressible flow over a circular cylinder.

Since VW-PINNs is meshfree, it cannot compute volume relying on mesh topology as mesh-based numerical methods. Therefore, based on kernel density estimation method [40-41], we develop a volume approximation algorithm suitable for random sampling, allowing for the approximate calculation of volume at extremely low time cost. Kernel density estimation is a non-parametric statistical method. Its fundamental concept involves placing a kernel function centered at each sampling point, such as the Gaussian kernel function utilized in this study. Subsequently, all these kernel functions are superimposed to estimate density. This method is widely applied in fields such as data analysis and machine learning.

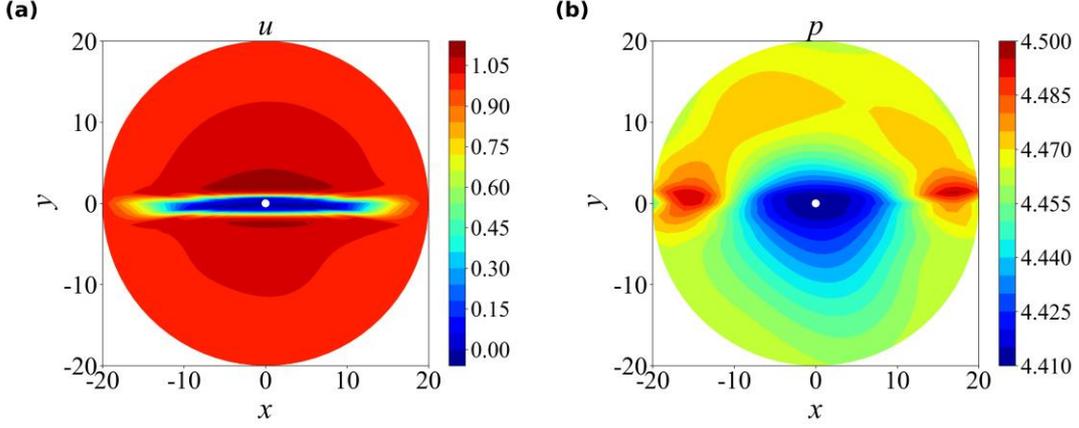

Figure 4. The flow field given by PINNs for solving inviscid compressible flow over a circular cylinder. (a) The $x$-component of the velocity field. (b) Pressure field.

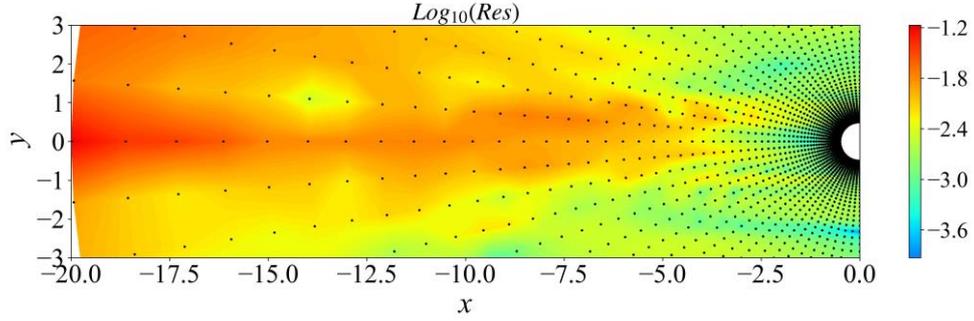

Figure 5. The residual distribution of Euler equations given by PINNs for solving inviscid compressible flow over a circular cylinder.

Before stating the volume approximation algorithm, it is important to note that PINNs solve the PDE in the entire spatiotemporal dimensions directly. Thus, we choose to incorporate the time coordinate into the volume calculation. Based on the collocation points coordinate $\{\mathbf{x}_r^i, t_r^i\}_{i=1}^{N_r}$, the probability density function $p(\mathbf{x},t)$ in the computational domain can be obtained:

$$p(\mathbf{x},t) = \frac{1}{N_r} \sum_{i=1}^{N_r} \exp(-\frac{\|\mathbf{x}-\mathbf{x}_r^i\|^2 + \|t-t_r^i\|^2}{2h^2}) \quad (10)$$

where $h > 0$ is the bandwidth, which controls the radial scope of the kernel functions. Based on formula (10), the probability density $p(\mathbf{x}_r^i, t_r^i)$ at each collocation point can be obtained. Calculate its reciprocal:

$$a(\mathbf{x}_r^i, t_r^i) = \frac{1}{p(\mathbf{x}_r^i, t_r^i)} \quad (11)$$

normalize $a(\mathbf{x}_r^i, t_r^i)$:

$$a_{norm}(\mathbf{x}_r^i, t_r^i) = \frac{a(\mathbf{x}_r^i, t_r^i)}{\sum_{i=1}^{N_r} a(\mathbf{x}_r^i, t_r^i)} \tag{12}$$

Finally, by multiplying $a_{norm}$ with the total volume $V_s$ of the entire computational domain, the approximate value of the volume occupied by each collocation point in the computational domain is obtained. A schematic illustration of VW-PINNs is shown in Figure 6.

$$V(\mathbf{x}_r^i, t_r^i) = V_s a_{norm}(\mathbf{x}_r^i, t_r^i) \tag{13}$$

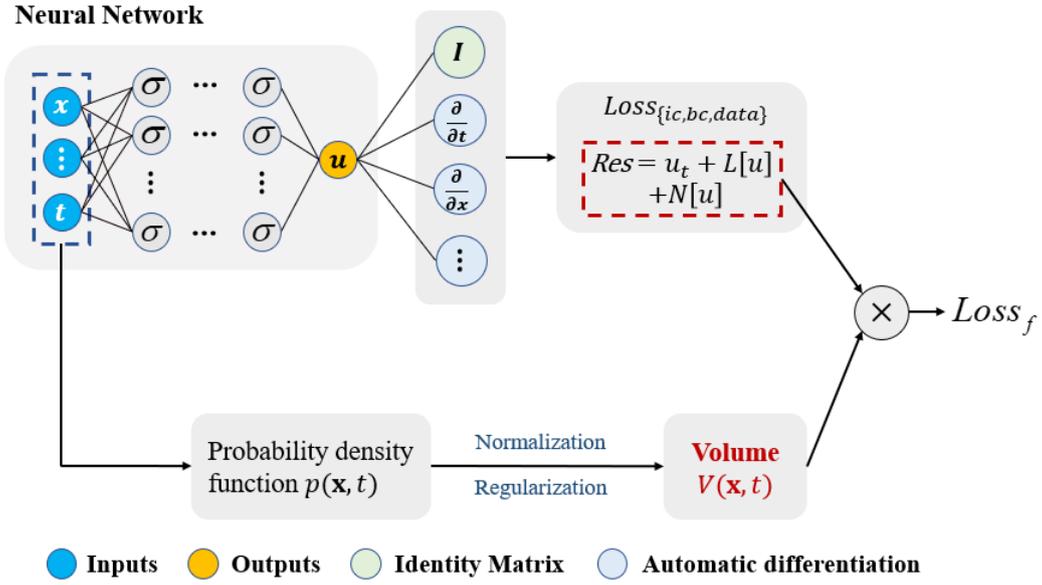

Figure 6. A Schematic of VW-PINNs for solving partial differential equations.

## 3 Results

We verify the effectiveness of VW-PINNs by solving four forward problems and one inverse problem. The forward problems include: inviscid compressible flow over a circular cylinder, viscous incompressible flow over a circular cylinder, viscous incompressible flow over the NACA0012 airfoil at 5 degrees angle of attack, and Burgers' equation based on adaptive sampling [30]. The inverse problem involves Burgers' equation. To evaluate the accuracy of the solution, we calculate the relative $L_2$ error:

$$\frac{\|U_\theta - U\|_2}{\|U\|_2} \tag{14}$$

where $U_\theta$ is the predicted solution by the neural network, $U$ represents the numerical solution. For three flow cases, $U$ is obtained through finite volume method, whereas for the Burgers' equation, we using spectral method to obtain numerical solution. The study develops programs based on the PyTorch platform and executes the algorithm on NVIDIA GeForce RTX 3080 GPU.

3.1 Inviscid compressible flow over a circular cylinder

We first validate the effect of VW-PINNs by solving the inviscid compressible flow over a circular cylinder as described in Section 2.2. The governing equations are formula (7). VW-PINNs maintains the same solving settings as PINNs.

Figure 7 shows the comparison of loss functions obtained by PINNs (Section 2.2) and VW-PINNs respectively. From the curves of the training loss, we observe that the boundary loss of both is reduced to the order of $10^{-6}$, indicating that the approximation error of the boundary conditions is adequately low. For the PDE loss of PINNs, in addition to exhibiting the conventional PDE loss that used to guide network optimization, we also calculate the volume-weighted PDE loss (In the optimization framework, as described in Section 2.3, the volume-weighted PDE loss better represents the convergence state of the total PDE residual in the entire computational domain). Obviously, the volume-weighted PDE loss of PINNs is only reduced by 3 orders of magnitude. Corresponding to the residual distribution (Figure 5), the total residual of the Euler equations doesn't sufficiently converge. In contrast, the volume-weighted PDE loss of VW-PINNs is reduced by 7 orders of magnitude, resulting in an efficient convergence of the total PDE residual. In Figure 8, we plot the flow field near the cylinder given by VW-PINNs, which are consistent with the physics. For the problem of flow over an object, the pressure coefficient on the object is usually of greater interest to researchers. Therefore, we use it to assess the accuracy of the solution, as shown in Figure 9. The relative $L_2$ error between the pressure coefficients obtained by VW-PINNs and the finite volume method is only 1.33%, while the result obtained by PINNs is obviously incorrect.

In Figure 10, we plot the distribution of equation residuals given by VW-PINNs. Compared with the result given by PINNs (Figure 5), our method achieves consistent and sufficient convergence of the residuals in the entire computational domain.

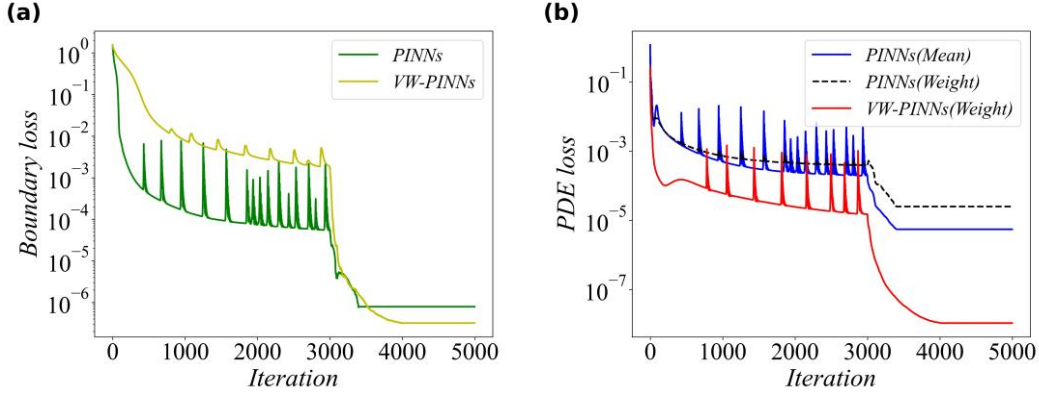

Figure 7. The comparison of loss functions obtained by PINNs (Section 2.2) and VW-PINNs for respectively solving inviscid compressible flow over a circular cylinder. (a) Boundary loss. (b) PDE loss. The blue solid line represents the conventional PDE loss (formula (5)) that used to guide PINNs optimization. The black dashed line represents the volume-weighted PDE loss (formula (9)) when PINNs are optimized based on conventional PDE loss. The red solid line represents the volume-weighted PDE loss that used to guide VW-PINNs optimization.

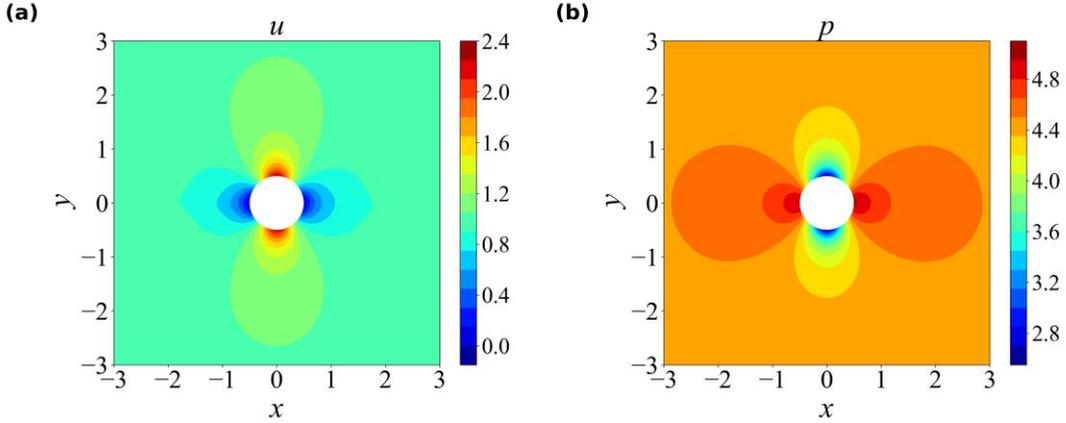

Figure 8. The flow field near the cylinder given by VW-PINNs for solving inviscid compressible flow over a circular cylinder. (a) The $x$-component of the velocity field. (b) Pressure field.

3.2 Viscous incompressible flow over a circular cylinder

Here we verify VW-PINNs by solving the viscous incompressible flow over the circular cylinder at the Reynolds number $Re = 40$. The dimensionless governing equations of this problem are:

$$\frac{\partial u}{\partial x} + \frac{\partial v}{\partial y} = 0$$
$$u\frac{\partial u}{\partial x} + v\frac{\partial u}{\partial y} + \frac{\partial p}{\partial x} - \frac{1}{Re}(\frac{\partial^2 u}{\partial x^2} + \frac{\partial^2 u}{\partial y^2}) = 0 \quad (15)$$
$$u\frac{\partial v}{\partial x} + v\frac{\partial v}{\partial y} + \frac{\partial p}{\partial y} - \frac{1}{Re}(\frac{\partial^2 v}{\partial x^2} + \frac{\partial^2 v}{\partial y^2}) = 0$$

where $u$ denotes the $x$-component of the velocity field, $v$ the $y$-component. $p$ represents pressure. When $Re = 40$, there exist two symmetric separated vortices behind the cylinder. We take $[x, y]$ as the network inputs and $[u, v, p]$ as the network outputs.

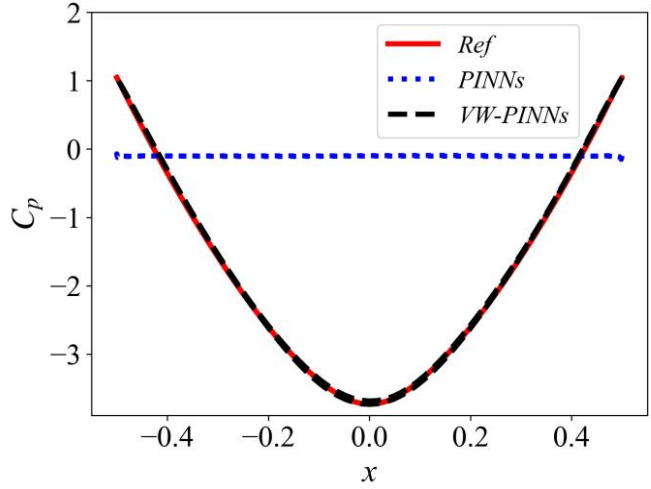

Figure 9. The comparison of pressure coefficients on the cylinder obtained by finite volume method (red), PINNs (blue) and VW-PINNs (black) for respectively solving inviscid compressible flow over a circular cylinder.

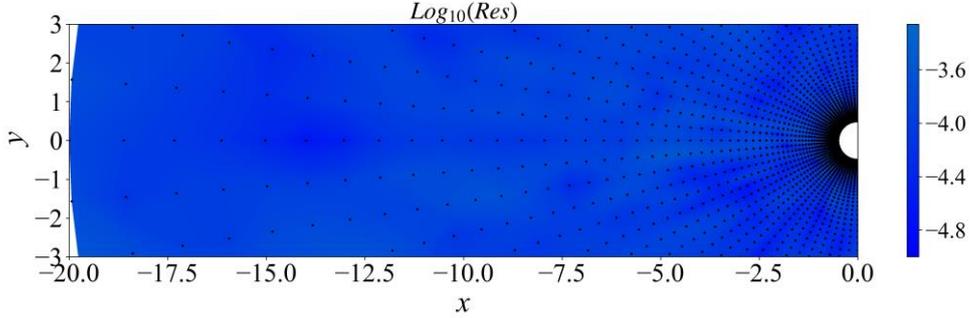

Figure 10. The residual distribution of Euler equations given by VW-PINNs for solving inviscid compressible flow over a circular cylinder.

The cylinder is placed at $(x, y) = (0, 0)$ with diameter $D = 1$. The computational domain and the distribution of collocation points are illustrated in Figure 11. The total number of collocation points is $N_r = 12300$. At the cylinder, inflow, and outflow boundaries, we randomly place $N_{bc} = 200$, 100, and 75 points respectively and impose no-slip condition $\mathbf{V} = \mathbf{0}$, velocity inlet condition $[u_\infty, v_\infty] = [1, 0]$, and pressure outlet condition $p_\infty = 0$. For the upper and lower boundaries, we randomly collect $N_{bc} = 100$ points and also apply velocity inlet condition. We solve the problem through PINNs and VW-PINNs respectively. The neural networks contain 5 hidden layers with 64 neurons per layer, and the activation function chosen as tanh. For the network training, we first run 3000 steps using the Adam optimizer, followed by an additional 2000 steps using the L-BFGS optimizer (maximum number of inner iterations is 20).

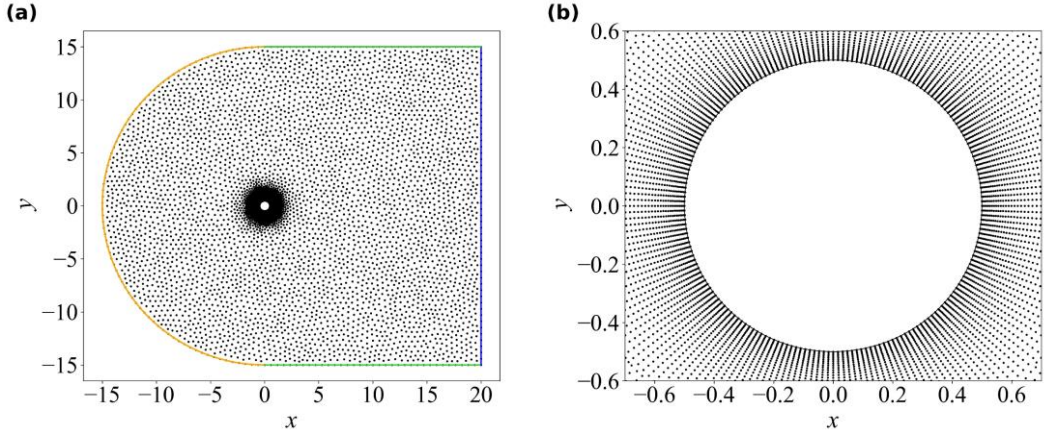

Figure 11. The computational domain and the distribution of collocation points in solving viscous incompressible flow over a circular cylinder. (a) Entire computational domain. The orange line represents the inflow boundary, the blue line represents the outflow boundary, and the green lines represent the upper and lower boundaries. (b) Near the cylinder region.

Figure 12 shows the flow fields given by PINNs and VW-PINNs respectively. Obviously, PINNs fail to solve, while the results given by VW-PINNs are reasonable. The convergence of loss functions obtained by two methods indicate the reason for their different results, as shown in Figure 13. When the boundary losses of both are low enough, the volume-weighted PDE loss of PINNs decreases by only two orders of magnitude, meaning that the total PDE residual in the entire computational domain doesn't sufficiently converge. By comparison, VW-PINNs reduce the volume-weighted PDE loss by 5 additional orders of magnitude. In Figure 14, we evaluate the accuracy of the solution by the pressure coefficient on the cylinder. Using the solution obtained by the finite volume method as a reference, the relative $L_2$ error of VW-PINNs is only 0.53%, whereas PINNs yield incorrect result.

In addition, we compare the streamline distribution near the cylinder given respectively through finite volume method and VW-PINNs in Figure 15. The two results exhibit strong consistency, indicating our method accurately obtains the location of separation points and the scale of separation vortices.

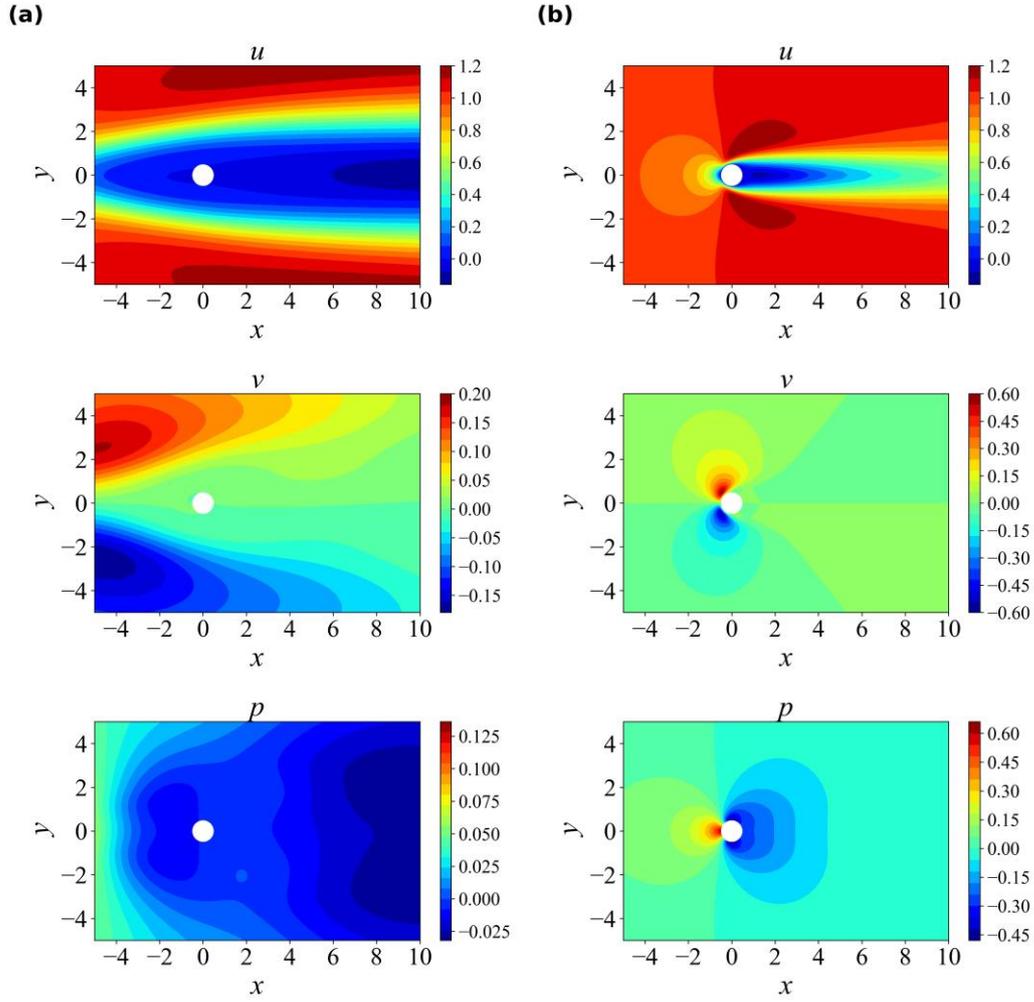

Figure 12. The comparison of flow fields near the cylinder given by (a) PINNs and (b) VW-PINNs for respectively solving viscous incompressible flow over a circular cylinder.

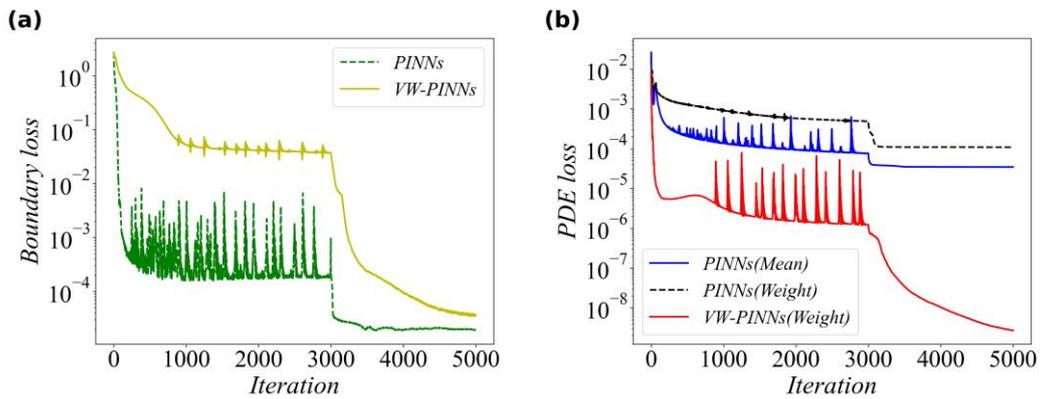

Figure 13. The comparison of loss functions obtained by PINNs and VW-PINNs for respectively solving viscous incompressible flow over a circular cylinder. (a) Boundary loss. (b) PDE loss.

For viscous incompressible flow, Rao et al. [45] suggested that using the constitutive form of the Navier-Stokes equations with a reduced order as the constraint in PINNs to enhance the accuracy. We employ this approach to reconfirm our perspective and evaluate the robustness of VW-PINNs. The constitutive form of the Navier-Stokes equations are:

$$\nabla \cdot \mathbf{V} = \mathbf{0}$$
$$(\mathbf{V} \cdot \nabla)\mathbf{V} = \nabla \cdot \boldsymbol{\sigma}$$
$$\boldsymbol{\sigma} = -p\mathbf{I} + \frac{1}{Re}(\nabla \mathbf{V} + \nabla \mathbf{V}^{\mathrm{T}}) \quad (16)$$
$$p = -tr\boldsymbol{\sigma}/2$$

where $\nabla$ is the Nabla operator, $\mathbf{V} = [u,v]$ is the velocity vector and $\boldsymbol{\sigma}$ is the Cauchy stress tensor (For two-dimensional problems, $\boldsymbol{\sigma}$ comprising three components). Consistent with [45], we define the stream function $\psi$, with $[x, y]$ as the network inputs and $[\psi, p, \boldsymbol{\sigma}]$ as the network outputs. $[u, v]$ can be calculated from $\psi$ under incompressible condition:

$$u = \frac{\partial \psi}{\partial y}, \quad v = -\frac{\partial \psi}{\partial x} \quad (17)$$

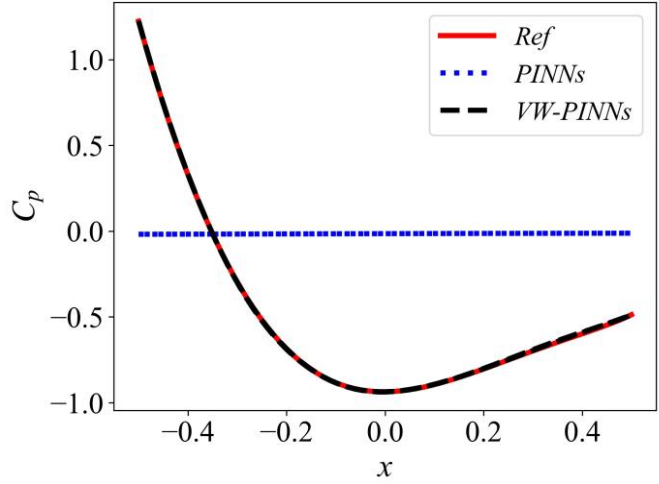

Figure 14. The comparison of pressure coefficients on the cylinder obtained by finite volume method (red), PINNs (blue) and VW-PINNs (black) for respectively solving viscous incompressible flow over a circular cylinder.

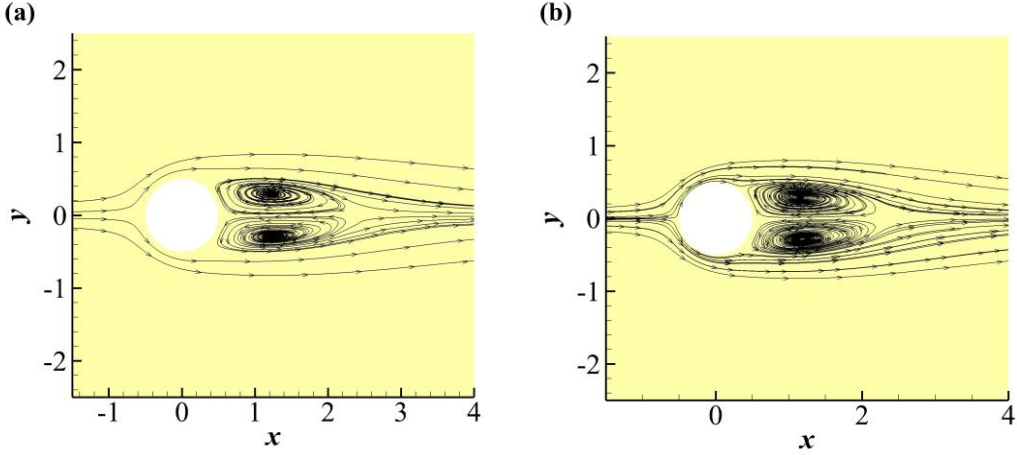

Figure 15. The comparison of streamline near the cylinder given by (a) finite volume method and (b) VW-PINNs for respectively solving viscous incompressible flow over a circular cylinder.

Except that the number of neurons in the network output layer increases from 3 to 5 (Due to $\sigma$ comprising three components), the solving settings are the same as those used in solving standard Navier-Stokes equations (formula (15)). We solve the problem through PINNs and VW-PINNs respectively. Figure 16 shows the comparison of pressure coefficients on the cylinder obtained by the two methods. Similar to Figure 14, the result obtained by VW-PINNs exhibits a high agreement with the numerical solution, whereas the result obtained by PINNs is notably incorrect. This is reasonable, as according to our analysis in Section 2.2, the equation form is not the primary factor causing the failure of solving the problems involving nonuniform collocation points.

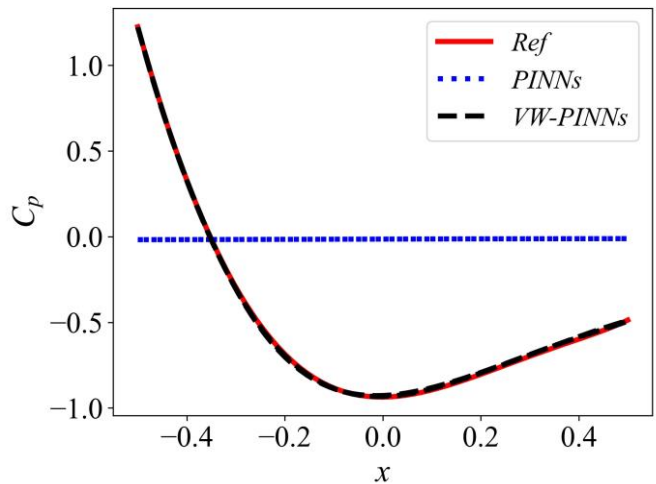

Figure 16. The comparison of pressure coefficients on the cylinder obtained by finite volume method (red), PINNs (blue) and VW-PINNs (black) for respectively solving the constitutive form of the Navier-Stokes equations.

## 3.3 Viscous incompressible flow over the NACA0012 airfoil

Our next example involves a scenario of the viscous incompressible flow over the NACA0012 airfoil. Compared with the cylinder, the curvature of the airfoil changes more rapidly. The governing equations of the problem are still formula (15). We consider the angle of attack to be 5 degree, $Re = 400$, with $[x, y]$ as the network inputs and $[u, v, p]$ as the network outputs.

The leading edge point of the airfoil is located at $(0,0)$ with chord $c = 1$. The computational domain and the distribution of collocation points are illustrated in Figure 17. The total number of collocation points is $N_r = 14640$. The setting of boundary conditions is the same as Section 3.2. We solve the problem through PINNs and VW-PINNs respectively. The neural networks contain 7 hidden layers with 64 neurons per layer, and the activation function chosen as tanh. For the network training, we first run 3000 steps using the Adam optimizer, followed by an additional 10000 steps using the L-BFGS optimizer (maximum number of inner iterations is 20).

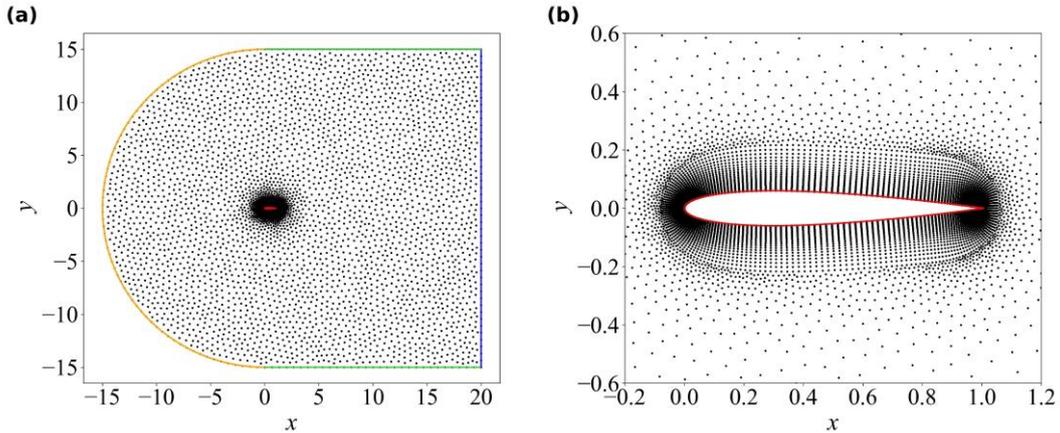

Figure 17. The computational domain and the distribution of collocation points in solving viscous incompressible flow over the NACA0012 airfoil. (a) Entire computational domain. The orange line represents the inflow boundary, the blue line represents the outflow boundary, the green lines represent the upper and lower boundaries, and the red line represents the airfoil. (b) Near the airfoil region.

Figure 18 shows the flow fields near the airfoil given by PINNs and VW-PINNs. Similar to Sections 3.1 and 3.2, VW-PINNs successfully solve viscous incompressible flow over the NACA0012 airfoil where PINNs fail. From the convergence of loss functions, we observe that compared with PINNs, our approach reduces the volume-weighted PDE loss by 3 additional orders of magnitude, as shown in Figure 19. In Figure 20, we still evaluate the accuracy of the solution through the pressure coefficient on the airfoil. The relative $L_2$ error between the result obtained by VW-PINNs and the numerical solution is 3.21%, while PINNs obtain a wrong result.

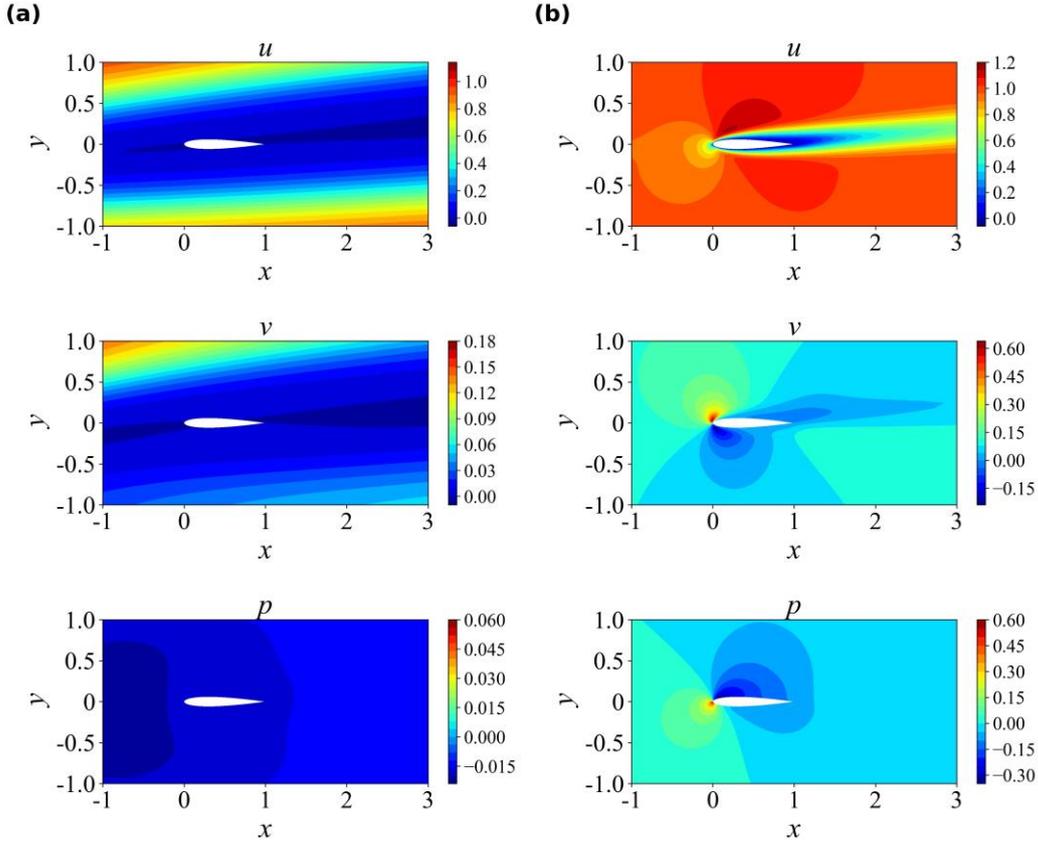

Figure 18. The comparison of flow fields near the airfoil given by (a) PINNs and (b) VW-PINNs for respectively solving viscous incompressible flow over the NACA0012 airfoil.

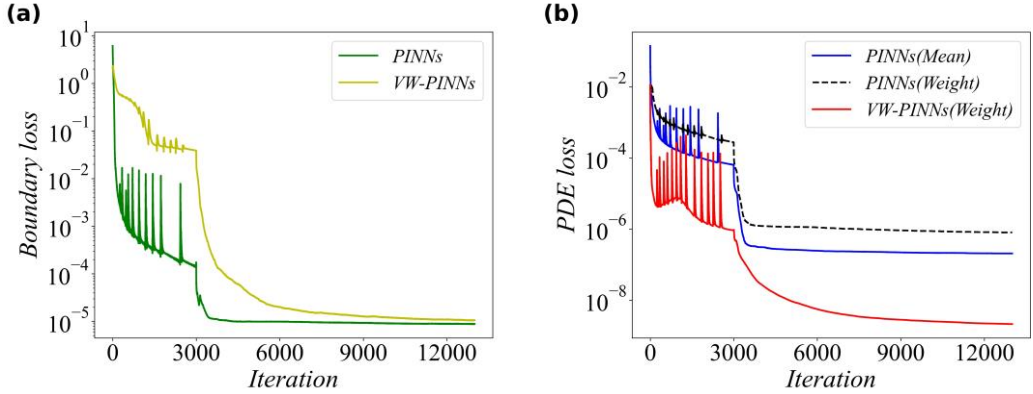

Figure 19. The comparison of loss functions obtained by PINNs and VW-PINNs for respectively solving viscous incompressible flow over the NACA0012 airfoil. (a) Boundary loss. (b) PDE loss.

3.4 Forward problem involving Burgers' equation based on adaptive sampling

In this section, we utilize the adaptive nonuniform sampling method based on the PINNs proposed by Wu et al. [30] to validate the effectiveness of VW-PINNs in adaptive scenarios.

In [30], the authors proposed an adaptive sampling method by evaluating PDE

residuals during network training. Specifically, after every predefined number of training steps, the PDE residuals $\varepsilon(\mathbf{x}_s, t_s)$ are calculated on a highly dense set of uniform sampling points $(\mathbf{x}_s, t_s)$ within the computational domain. Then, the probability density function $p(\mathbf{x}_s, t_s)$ for adaptive sampling can be calculated:

$$p(\mathbf{x}_s, t_s) \propto \frac{\varepsilon^k(\mathbf{x}_s, t_s)}{\mathrm{E}[\varepsilon^k(\mathbf{x}_s, t_s)]} + c \tag{18}$$

where E[·] represents the mean operator. $k$ and $c$ are two hyperparameters, which determine the distribution of sampling points. According to the authors' descriptions, several other adaptive sampling methods [32-39] can all be summarized by formula (18), differing only in the values taken by $k$ and $c$. Compared with uniform sampling, this method significantly enhances the solution accuracy of PINNs with sparse collocation points, especially when the solution of PDE has steep gradients.

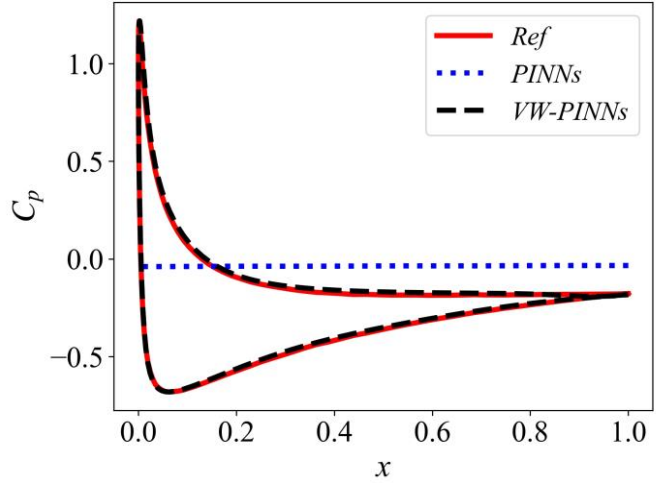

Figure 20. The comparison of pressure coefficients on the NACA0012 airfoil obtained by finite volume method (red), PINNs (blue) and VW-PINNs (black) for respectively solving viscous incompressible flow over the NACA0012 airfoil.

We conduct the testing by solving Burgers' equation (formula (19)) in [30].

$$\begin{aligned}
&\frac{\partial u}{\partial t} + u\frac{\partial u}{\partial x} = v\frac{\partial^2 u}{\partial x^2}, \quad x \in [-1,1], t \in [0,1] \\
&u(x,0) = -\sin(\pi x) \\
&u(-1,t) = u(1,t) = 0
\end{aligned} \tag{19}$$

where $u$ is the flow velocity and $v = 0.01/\pi$ is the viscosity of the fluid. The solution of the Burgers' equation has a sharp front when $x = 0$ and $t$ is close to 1, as shown in Figure 21. We integrate algorithm based on the author's publicly available source code and adopt the solving settings as specified in the article. The number of collocation points is fixed at $N_r = 2000$, and hard constraints are imposed on the definite conditions. The neural networks contain 4 hidden layers with 64 neurons per layer, and the the activation function chosen as tanh. When solving the equation, firstly, based on uniform sampling, the neural network is trained using 15000 steps of Adam and then 1000 steps of L-BFGS (maximum number of inner iterations is 1). Subsequently, the adaptive sampling is employed (we use the best hyperparameter setting $k = 1, c = 1$ from the original literature). Within a sampling period, the neural network is trained using 1000 steps of Adam and then 1000 steps of L-BFGS.

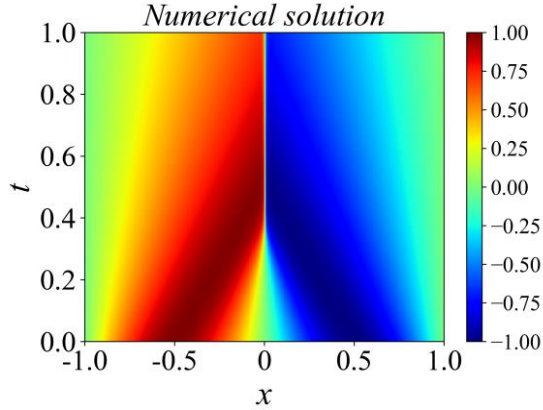

Figure 21. The numerical solution of Burgers' equation

Figure 22 shows the relative $L_2$ error obtained by the original method and VW-PINNs. The region I represents the training process based on uniform sampling, whereas region II corresponds to the adaptive sampling process. The result given by the original method is in high agreement with the result in [30]. When the volume weighting method is integrated, it is obvious that the solving of the new framework has a faster convergence speed, with a speedup of 2.22. If only the adaptive sampling training process is considered, the speedup reaches 3.2.

3.5 Inverse problem involving Burgers' equation

The solving of inverse problems involving PDE has been a major limitation of numerical methods [6]. PINNs can conveniently incorporate various observed data and achieve the solving of such problems at an extremely low cost. Therefore, we finally validate the VW-PINNs by solving the inverse problem involving Burgers' equation.

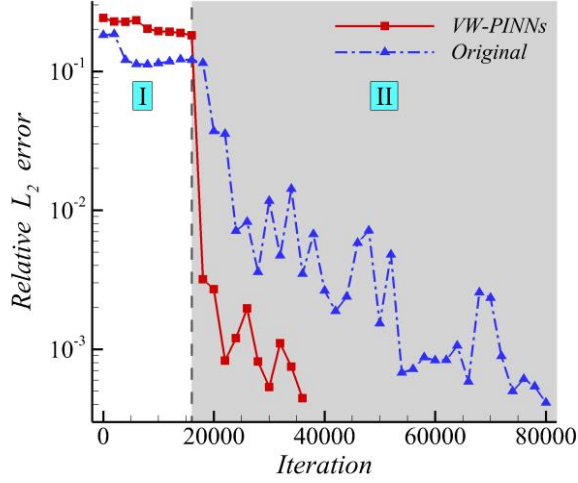

Figure 22. The relative $L_2$ error obtained by original method (blue) and VW-PINNs (red) for respectively solving Burgers' equation.

The governing equation of this problem is:

$$\frac{\partial u}{\partial t} + \lambda_1 u \frac{\partial u}{\partial x} = \lambda_2 \frac{\partial^2 u}{\partial x^2}, \quad x \in [-1,1], t \in [0,1] \tag{20}$$

where $\lambda_1$ and $\lambda_2$ are the unknown parameters, and their exact values are $\lambda_1 = 1, \lambda_2 = 0.01/\pi$. The known conditions of the problem consist of only 200 randomly observed data in the computational domain, all other informations (including initial condition and boundary condition) are unknown. The goal of solving is to obtain a high-resolution $u$ while identifying the parameters $\lambda_1$ and $\lambda_2$.

In addition to the observed data, we randomly sample $N_r = 1000$ collocation points. According to $u$ having a steep gradient at $x = 0$, we perform local refinement in this region (sampling density is 5 times that of other regions), as shown in Figure 23. The network architecture follows the setting from [1], which contains 8 hidden layers with 20 neurons per layer, and the activation function chosen as tanh. For the network training, we first run 2000 steps using the Adam optimizer, followed by an additional 6000 steps using the L-BFGS optimizer (maximum number of inner iterations is 1).

Figure 24 shows the relative errors of $\lambda_1, \lambda_2$ and $u$ obtained by PINNs and VW-PINNs. We observe that VW-PINNs solutions are more accurate, especially in identifying the viscosity coefficient $\lambda_2$ with a relative error of only 1.48%, while PINNs yield an error of 19.82%. With more accurate equation identification, it is natural that our method obtains a more precise $u$. These results demonstrate the ability of VW-PINNs to solve inverse problems under nonuniform sampling.

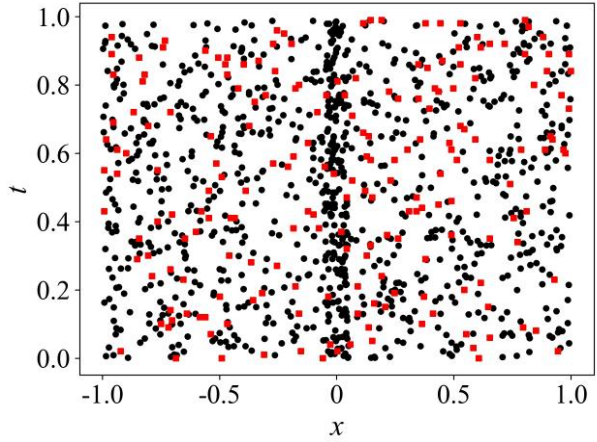

Figure 23. The distribution of sampling points in solving the inverse problem involving Burgers' equation. Black circles represent collocation points. Red squares represent observed data.

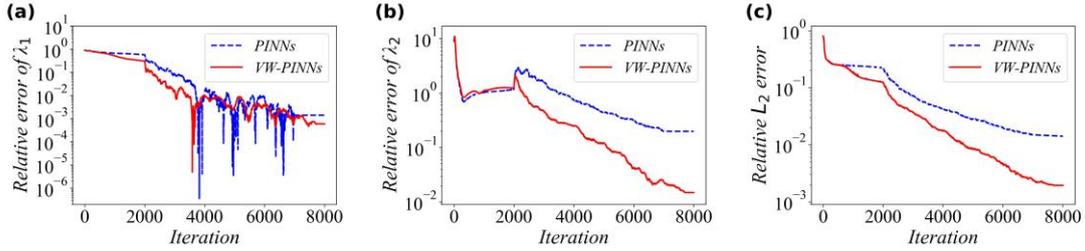

Figure 24. The relative errors in solving the inverse problem involving Burgers' equation. (a) $\lambda_1$. (b) $\lambda_2$. (c) Evaluating $u$ at a set of dense test points (the number is $2.56 \times 10^4$)

## 4 Conclusions

In this work, we first study the ill-conditioning of the PDE loss function for PINNs under nonuniform collocation points. Since the proportion of PDE residuals at all sampling points in the PDE loss is equal, when employing nonuniform sampling, the loss reduction benefits brought by reducing the local residuals in high-density regions are higher than those in low-density regions. Inevitably, the network will naturally focus on reducing residuals in high-density regions. In contrast, other low-density regions receive little attention and the residuals converge difficultly, resulting in the poor performance of PINNs. By solving the inviscid compressible flow over a circular cylinder, we analyze this problem specifically.

To address this limitation of PINNs, we define the volume-weighted residual and propose volume-weighted physics-informed neural networks (VW-PINNs). Through weighting the PDE residuals by the volume that the collocation points occupy within the computational domain, we embed explicitly the spatial distribution characteristics of collocation points in the residual evaluation. Relying on the linear inverse relationship between the sampling density of collocation points and the volume they occupy, VW-PINNs balance the differences in residual convergence across regions with various sampling densities in network optimization. The decrease of the volume-weighted PDE loss ensures efficient convergence of the total PDE residual across the entire computational domain. To calculate the volume of collocation points, we also develop a volume approximation algorithm suitable for random sampling based on the kernel density estimation method. It only requires the coordinates of sampling points to calculate the approximate volume. By solving four forward problems and one inverse problem, we verify the effectiveness of the proposed method:

(1) VW-PINNs successfully solve the flow over a circular cylinder and the flow over the NACA0012 airfoil under different inflow conditions where conventional PINNs fail, with the relative errors of wall pressure coefficient only about 1% and accurate capture of the flow separation behind the cylinder.

(2) By incorporating the volume weighting method into the existing adaptive sampling approach, we triple the solving efficiency of the adaptive sampling approach.

(3) In solving the inverse PDE problem with nonuniform sampling, compared with PINNs, VW-PINNs reduce the relative error by an order of magnitude.

However, our method still faces challenges in solving flows at high Reynolds numbers and flow over the airfoil under compressible condition, which is possibly due to the lack of fitting ability of neural networks for flow with significant scale differences. These challenges may be solved in the future by using recent advances in the field.

## Acknowledgements

We would like to acknowledge the support of the National Natural Science Foundation of China (No. 92152301).